\documentclass[journal=jctcce,manuscript=article]{achemso}
\usepackage{achemso}

\usepackage{amsmath, amssymb}
\usepackage{color}
\usepackage{latexsym}
\usepackage{indentfirst}
\usepackage{graphicx}
\usepackage{placeins}
\usepackage{booktabs}
\usepackage{algorithm}
\usepackage{algorithmic}
\usepackage{multirow}
\usepackage{subfig}
\newcommand{\ud}{\,\mathrm{d}}

\newcommand{\Or}{\mathcal{O}}

\newcommand{\wt}[1]{\widetilde{#1}}

\newcommand{\abs}[1]{\left\lvert#1\right\rvert}
\newcommand{\norm}[1]{\left\lVert#1\right\rVert}

\newcommand{\vace}{V_{X}^{\mathrm{ACE}}}

\newcommand{\barint}{\kern4pt \raise3.4pt\hbox{\vrule height.6pt
    width7pt} \kern-11pt \int}

\newcommand{\bvec}[1]{\mathbf{#1}}
\newcommand{\Vion}{V_{\mathrm{ion}}}

\newcommand{\vr}{\bvec{r}}

\renewcommand{\eqref}[1]{(\ref{#1})}

\title{Adaptively Compressed Exchange Operator}
\author{Lin Lin}
\email{linlin@math.berkeley.edu}
\affiliation[University of California, Berkeley]{
Department of Mathematics, University of California, Berkeley, CA 94720, USA}
\alsoaffiliation[Lawrence Berkeley National Laboratory]{
Computational Research Division, Lawrence Berkeley National Laboratory,
Berkeley, CA 94720, USA}

\date{\today}

\begin{document}

\begin{abstract}
The Fock exchange operator plays a central role in modern quantum
chemistry. The large computational cost associated with the Fock
exchange operator hinders Hartree-Fock calculations and Kohn-Sham
density functional theory calculations with hybrid exchange-correlation
functionals, even for systems consisting of hundreds of atoms. We
develop the adaptively compressed exchange operator (ACE) formulation,
which greatly reduces the computational cost associated with the Fock
exchange operator without loss of accuracy. The ACE formulation does not
depend on the size of the band gap, and thus can be applied to insulating,
semiconducting as well as metallic systems. In an iterative framework
for solving Hartree-Fock-like systems, the ACE formulation only requires
moderate modification of the code, and can be potentially beneficial for
all electronic structure software packages involving exchange
calculations. Numerical results indicate that the ACE formulation can
become advantageous even for small systems with tens of atoms. In particular,
the cost of each self-consistent field iteration for the electron density in the ACE formulation is only marginally larger than that of the generalized gradient approximation (GGA) calculation, and thus offers orders of magnitude speedup for Hartree-Fock-like calculations.
\end{abstract}

\section{Introduction}\label{sec:intro}

The Fock exchange operator, or simply the exchange operator, plays a
central role both in wavefunction theory and in density functional
theory, two cornerstones of modern quantum chemistry. Hartree-Fock
theory (HF) is the starting point of nearly all wavefunction based
correlation methods. Kohn-Sham density functional theory
(KSDFT)~\cite{HohenbergKohn1964,KohnSham1965} is the most widely used
electronic structure theory for molecules and systems in condensed
phase. The accuracy of KSDFT is
ultimately determined by the exchange-correlation (XC) functional
employed in the calculation. Despite the great success of relatively
simple XC functionals such as local density approximation
(LDA)~\cite{CeperleyAlder1980,PerdewZunger1981},
generalized gradient approximation
(GGA)~\cite{Becke1988,LeeYangParr1988,PerdewBurkeErnzerhof1996} and
meta-GGA~\cite{StaroverovScuseriaTaoEtAl2003,ZhaoTruhlar2008}
functionals, numerous computational studies in the past two decades
suggest that KSDFT calculations with hybrid
functionals~\cite{Becke1993,PerdewErnzerhofBurke1996,HeydScuseriaErnzerhof2003,HeydScuseriaErnzerhof2006}
can provide systematically
improved description of important physical quantities
such as band gaps, for a
vast range of systems. As an example, the B3LYP functional~\cite{Becke1993}, which is only one specific hybrid
functional, has
generated more than $55,000$ citations (Data from ISI Web of Science,
January, 2016).
Hybrid functional calculations are computationally more involved since it
contains a fraction of the Fock exchange
term, which is defined using the entire density matrix rather than the
electron density. If the exchange operator is constructed explicitly,
the computational cost scales as $\Or(N_e^4)$ where $N_e$ is the number
of electrons of the system. The cost can be reduced to $\Or(N_e^3)$ by
iterative algorithms that avoid the explicit construction of the
exchange operator, but with a large preconstant.  Hence hybrid
functional calculations for systems consisting of hundreds of atoms or
even less can be a very challenging computational task.


Various numerical methods have been developed to reduce the
computational cost of Hartree-Fock-like calculations (i.e. Hartree-Fock
calculations and KSDFT calculations with hybrid
functionals), most notably methods with asymptotic ``linear scaling''
complexity~\cite{Goedecker1999,BowlerMiyazaki2012}. The linear scaling methods use the fact that for
an insulating system with a finite HOMO-LUMO gap, the subspace spanned by
the occupied orbitals has a compressed representation: it is possible to
find a unitary transformation to transform all occupied orbitals into a
set of orbitals localized in the real space. This is closely related to 
the ``nearsightedness'' of electronic
matters~\cite{Kohn1996,ProdanKohn2005}. Various efforts have been
developed to find such localized
representation~\cite{FosterBoys1960,MarzariVanderbilt1997,WannierReview,Gygi2009,ELiLu2010,OzolinsLaiCaflischEtAl2013,DamleLinYing2015}.
After such localized representation is obtained, the exchange operator
also becomes simplified, leading to more efficient
numerical
schemes for systems of sufficiently large sizes~\cite{WuSelloniCar2009,GygiDuchemin2012,DamleLinYing2015}.
Recent numerical studies indicate that linear scaling methods can be
very successful in reducing the cost of the calculation of the exchange
term for systems of large sizes with substantial band gaps~\cite{ChenWuCar2010,DiStasioSantraLiEtAl2014,DawsonGygi2015}.

In this work, we develop a new method for reducing the computational
cost due to the Fock exchange operator. Our method aims at finding a low rank
decomposition of the exchange operator. However, standard low rank
decomposition schemes such as the singular value decomposition mandates the
low rank operator to yield similar result as the exchange
operator does when applied to an arbitrary orbital. This is doomed to
fail since the exchange operator is not a low rank operator, and
forcefully applied low rank decomposition can lead to unphysical
results. The key observation of this work is that in order to compute
physical quantities in Hartree-Fock-like calculations, it is sufficient to construct an  operator that
yields the same result as the exchange operator does when applied to the
\textit{occupied orbitals}. This is possible since the rank of the subspace spanned
by the occupied orbitals is known \textit{a priori}.
Since occupied orbitals vary in self-consistent field
iterations, the compressed representation must be adaptive to the
changing orbitals. Hence our compressed exchange operator
is referred to as the adaptively compressed exchange operator (ACE). 

The ACE formulation has a few notable advantages:  1) The ACE is a strictly
low rank operator, and there is no loss of accuracy when used to compute physical
quantities such as total energies and band gaps. 
2) The effectiveness of the ACE \textit{does not} depend on the size of the band
gap. Hence the method is applicable to insulators as well as
semiconductors or even metals. 3) The construction cost of the ACE is
similar to the one time application cost of the exchange operator to the set of occupied
orbitals. Once constructed, the ACE can be repeatedly
used. The cost of applying the ACE is similar to that of applying a
nonlocal pseudopotential operator, thanks to the low rank structure. 4)
In an iterative framework for solving the Hartree-Fock-like equations,
the ACE formulation only requires moderate change of the code, and 
could be potentially beneficial for all electronic structure software
packages involving exchange calculations. 


Our numerical results indicate that once the ACE is constructed, the
cost of each self-consistent field iteration (SCF) of the electron density
in a hybrid functional calculation is only marginally larger than that
of a GGA calculation.  The ACE formulation offers significant speedup
even for small systems with tens of atoms in a serial implementation.
For moderately larger systems, such as a silicon system with $216$ atoms, we
observe  more than $50$ times speedup in terms of the cost
of each SCF iteration for the electron density.

The rest of paper is organized as follows. Section~\ref{sec:prelim}
reviews the basic procedure of using iterative methods to solve
Hartree-Fock-like equations.
Section~\ref{sec:ace} describes the method of adaptively compressed
exchange operator. The numerical results are presented in
section~\ref{sec:numer}, followed by conclusion and future work in
section~\ref{sec:conclusion}.

\section{Iterative methods for solving Hartree-Fock-like equations}\label{sec:prelim}

For the sake of simplicity, our discussion below focuses on the
Hartree-Fock (HF) equations. The generalization from HF equations to KSDFT
equations with hybrid functionals is straightforward, and will be
mentioned at the end of this section.
To simplify notation we neglect the spin degeneracy in the discussion
below and assume all orbitals $\{{\psi}_{i}(\vr)\}$ are real.  The spin
degeneracy is properly included in the numerical results in
section~\ref{sec:numer}.

The HF theory requires solving the
following set of equations in a self-consistent fashion.
\begin{equation}
  \begin{split}
    &H\left[\{\psi_{j}\}\right]\psi_{i} = \left(-\frac12 \Delta  +
    \Vion +
    V_{H}[\rho] + V_{X}[\{\psi_{j}\}]\right)\psi_{i} = \varepsilon_{i} {\psi}_{i},\\
    &\int {\psi}^{*}_{i}(\vr) {\psi}_{j}(\vr) \ud \vr = \delta_{ij},
    \quad
    {\rho}(\vr) = \sum_{i=1}^{N_e} \abs{{\psi}_{i}(\vr)}^2.
  \end{split}
  \label{eqn:HF}
\end{equation}
Here the eigenvalues $\{\varepsilon_{i}\}$ are ordered non-decreasingly,
and $N_e$ is the number of electrons.  $\Vion$ is a local potential
characterizing the electron-ion interaction in all-electron
calculations. In pseudopotential
or effective core potential calculations, $\Vion$ may contain a low rank and semi-local component as well. $\Vion$ is independent of the
electronic states $\{\psi_{i}\}$. 
$\rho(\vr)$ is the electron
density. The Hartree potential is a local potential, and depends
only on the electron density as
\[
{V}_{H}[{\rho}](\vr,\vr')= \delta(\vr-\vr')\int \frac{\rho(\vr')}{\abs{\vr-\vr'}} \ud
\vr' .
\]
The exchange operator $V_{X}$ is a full rank, nonlocal operator, and depends on not only the density but also the occupied orbitals 
$\{\psi_{j}\}_{j=1}^{N_{e}}$ as
\begin{equation}
  V_{X}[\{\psi_{j}\}](\vr,\vr') = 
  -\sum_{j=1}^{N_{e}} 
  \frac{\psi_{j}(\vr) \psi_{j}(\vr')}{|\vr-\vr'|} \equiv
  -\frac{\Gamma(\vr,\vr';\{\psi_{j}\})}{\abs{\vr-\vr'}}.
  \label{eqn:VXkernel}
\end{equation}
Here $\Gamma(\vr,\vr';\{\psi_{j}\})=\sum_{j=1}^{N_{e}} \psi_{j}(\vr) \psi_{j}(\vr')$
is the single particle density matrix with an exact rank $N_e$.
However, $V_X$ is not a low rank operator due to the dot product (i.e. the Hadamard product) between $\Gamma$ and the Coulomb kernel.
One common way to solve the HF equations~\eqref{eqn:HF} is to expand the
orbitals $\{\psi_{j}\}_{j=1}^{N_{e}}$ using a small basis set
$\{\chi_{\mu}\}_{\mu=1}^{N_{\mu}}$, such as Gaussian type orbitals, Slater type
orbitals and numerical atomic orbitals.
The basis set is small in the sense
that the ratio $N_{\mu}/N_{e}$ is is a small constant (usually in the order of $10$).  This results in a Hamiltonian
matrix $H$ with reduced dimension $N_{\mu}$. In order to compute the
matrix element of $H$, the four-center integral 
\[
\iint\frac{\chi_{\mu}(\vr)\chi_{\alpha}(\vr)\chi_{\beta}(\vr')\chi_{\nu}(\vr')}{|\vr-\vr'|}
\ud \vr \ud \vr', \quad \alpha,\beta,\mu,\nu=1,\ldots,N_{\mu}
\]
needs to be performed. The cost of the four-center integral is
$\Or(N_{\mu}^{4})$. The quartic scaling becomes very expensive for systems of
large sizes.

For a more complete basis set such as planewaves and finite elements, the
constant $N_{\mu}/N_{e}$ is much larger (usually $1000$ or more), and the cost of forming all
four-center integrals is prohibitively expensive even for very small
systems. In such case, it is only viable to use an iterative algorithm,
which only requires the application of $V_X$ to a number of orbitals, rather than the explicit construction of $V_X$. According to
Eq.~\eqref{eqn:VXkernel}, $V_{X}$ applied to any orbital $\psi$ can be
computed as
\begin{equation}
  \left(V_{X}[\{\psi_{j}\}]\psi\right)(\vr) = 
  -\sum_{j=1}^{N_{e}} \psi_{j}(\vr) \int
  \frac{ \psi_{j}(\vr')\psi(\vr')}{|\vr-\vr'|} \ud \vr'.
  \label{eqn:applyVX}
\end{equation}
Eq.~\eqref{eqn:applyVX} can be performed by solving $N_{e}$ Poisson type
problems with an effective charge of the form $\psi_{j}(\vr')\psi(\vr')$. For instance,
in planewave calculations, if we denote by $N_{g}\equiv N_{\mu}$ the
total number of planewaves, then the cost for solving each Poisson
equation is
$\Or(N_{g}\log N_{g})$ thanks to techniques such as the Fast Fourier Transform (FFT). Applying $V_{X}$ to all $\psi_{i}$'s 
requires the solution of $N_{e}^2$ Poisson problems, and the total cost is
$\Or(N_{g}\log (N_{g})N_{e}^2)$.  The cubic scaling makes 
iterative algorithms asymptotically less expensive
compared to  quartic scaling algorithms associated with the four-center integral
calculation. Therefore for large systems, iterative methods 
can become attractive even for calculations  with
small basis sets.

The HF equations need to be performed self-consistently until the output
orbitals $\{\psi_{j}\}_{j=1}^{N_{e}}$ from Eq.~\eqref{eqn:HF} agree with those provided as the input to the Hamiltonian operator. However, the Fock exchange energy is only a small fraction (usually less than $5\%$) of the total energy, and it is more efficient \textit{not} to update the exchange operator in each self-consistent field iteration. For instance, in planewave based electronic structure software packages such as Quantum ESPRESSO~\cite{GiannozziBaroniBoniniEtAl2009}, the 
self-consistent field (SCF) iteration of all occupied orbitals can be separated into two
sets of SCF iterations. In the inner SCF iteration, the orbitals
defining the exchange operator $V_X$ as in Eq.~\eqref{eqn:VXkernel} are
fixed, denoted by $\{\varphi_{i}\}_{i=1}^{N_{e}}$. 
Then the matrix-vector multiplication of $V_{X}$ and an orbital $\psi$ is given by
\begin{equation}
  \left(V_{X}[\{\varphi_{j}\}]\psi\right)(\vr) = 
  -\sum_{j=1}^{N_{e}} \varphi_{j}(\vr) \int
  \frac{ \varphi_{j}(\vr')\psi(\vr')}{|\vr-\vr'|} \ud \vr'.
  \label{eqn:applyVX2}
\end{equation}
With $V_X$ fixed, the Hamiltonian operator only depends on the electron
density $\rho$, which needs to
be updated in the inner SCF iteration. This allows standard charge mixing
schemes, such as Anderson acceleration~\cite{Anderson1965} and Pulay
mixing~\cite{Pulay1980} to be used to converge the electron density efficiently. Note
that similar techniques to mix the density matrix directly can be
prohibitively expensive for large basis sets. 
Once the inner SCF for the electron density is converged, the
output orbitals can simply then be used as the input orbitals to update
the exchange operator. The outer SCF iteration continues until
convergence is reached. The convergence of the outer iteration can be
monitored by the convergence of the Fock exchange energy, defined as
\begin{equation}
  \label{eqn:EX}
  E_{X}^{HF} = -\frac{1}{2} \sum_{i,j=1}^{N_e} \iint
  \psi_{i}(\vr)\psi_{j}(\vr) \psi_{j}(\vr')
  \psi_{i}(\vr')\frac{1}{|\vr-\vr'|} \ud \vr \ud \vr'. 
\end{equation}

In each inner SCF iteration, with both $\rho$ and $\varphi_i$'s fixed, the Hamiltonian operator $H$ becomes a linear operator, and the linear eigenvalue problem
\begin{equation}
   (-\frac12 \Delta  + \Vion + V_{H}[\rho] + V_{X}[
   \{\varphi_{j}\}])\psi_{i} = \varepsilon_i \psi_i
  \label{eqn:HFlineig}
\end{equation}
needs to be solved. The linear eigenvalue problem can be solved 
with iterative algorithms such as the Davidson
method~\cite{Davidson1975} and the
locally optimal block preconditioned conjugated gradient (LOBPCG)
method~\cite{Knyazev2001}.  Alg.~\ref{alg:fock} describes the pseudocode
of using iterative methods to solve Hartree-Fock-like equations.

\begin{algorithm}[htbp]
\begin{center}
  \begin{minipage}{5in}
\begin{algorithmic}[1]
  \WHILE{exchange energy is not converged}
  \WHILE{electron density $\rho$ is not converged}
  \STATE Solve the linear eigenvalue problem~\eqref{eqn:HFlineig} with iterative schemes.
  \STATE Update $\rho^{out}(\vr)\gets \sum_{i=1}^{N_e}
  \abs{{\psi}_{i}(\vr)}^2$.
  \STATE Update $\rho$ using $\rho^{out}$ and possibly previous history of $\rho$ with charge mixing schemes.
  \ENDWHILE
  \STATE Compute the exchange energy $E_{X}$.
  \STATE Update $\{\varphi_{j}\}_{j=1}^{N_{e}}\gets
  \{\psi_{j}\}_{j=1}^{N_{e}}$.
  \ENDWHILE
\end{algorithmic}
\end{minipage}
\end{center}
\caption{Iterative methods for solving Hartree-Fock-like
equations.}
\label{alg:fock}
\end{algorithm}

So far our discussion focuses on the Hartree-Fock theory. For KSDFT calculations with hybrid functionals, such as the PBE0 functional~\cite{PerdewErnzerhofBurke1996}, the exchange-correlation energy is
\begin{equation}
E_{xc}^{PBE0} = \frac{1}{4} E_X^{HF} + \frac{3}{4} E_X^{PBE} + E_{c}^{PBE}. 
\label{eqn:ExPBE0}
\end{equation}
Here $E_X^{PBE}$ and $E_{c}^{PBE}$ are the exchange and correlation part of the energy from the GGA-type  
Perdew-Burke-Ernzerhof (PBE) functional~\cite{PerdewBurkeErnzerhof1996},
respectively. Hence the corresponding exchange operator $V_X^{PBE0}$ is
simply given by $1/4$ of the exchange operator defined in
Eq.~\eqref{eqn:VXkernel}. For exchange-correlation functionals with screened exchange interactions such as the HSE functional~\cite{HeydScuseriaErnzerhof2003}, the exchange-correlation energy is
\begin{equation}
E_{xc}^{HSE}(\mu) = \frac{1}{4} E_X^{SR}(\mu) + \frac{3}{4} E_X^{PBE,SR}(\mu) + E_X^{PBE,LR} + E_{c}^{PBE}. 
\label{eqn:ExHSE}
\end{equation}
Here $E_X^{PBE,SR}$ and $E_X^{PBE,LR}$ refers to short range and long range part of the exchange contribution in the PBE functional, respectively. $E_X^{SR}$ is the short range part of the Fock exchange energy, defined as
\begin{equation}
  \label{eqn:EXSR}
  E_{X}^{SR}(\mu) = -\frac{1}{2} \sum_{i,j=1}^{N_e} \iint
  \psi_{i}(\vr)\psi_{j}(\vr) \psi_{j}(\vr')
  \psi_{i}(\vr')\frac{\mathrm{erfc}(\mu(|\vr-\vr'|))}{|\vr-\vr'|} \ud \vr \ud \vr'. 
\end{equation}
Here $\mathrm{erfc}$ is the complementary error function, and $\mu$ is
an adjustable parameter to control the screening length of the short
range part of the Fock exchange interaction. The only change is to
replace the Coulomb kernel by the screened Coulomb kernel, and the screened Coulomb kernel should be used
to define the exchange operator $V_X^{HSE}$ accordingly.

\section{Adaptively compressed exchange operator}\label{sec:ace}

The most expensive step of Alg.~\ref{alg:fock} is the matrix-vector
multiplication between the Fock operator $V_X$ and all occupied
orbitals. Each set of such matrix-vector multiplication amounts to the
solution of $N_{e}^2$ Poisson equations. This needs to be done for each
iteration step when solving the linear eigenvalue problem~\eqref{eqn:HFlineig},
and in each inner SCF iteration for updating the electron density.

In order to reduce the computational cost, it is desirable to use a low rank decomposition to approximate the Fock exchange operator $V_X$. 
However, the exchange operator is a full rank operator, and a compressed
representation, such as the singular value decomposition (SVD), can
lead to inaccurate results.  However, note that the goal of a singular value decomposition of is to find
an effective operator, denoted by $\wt{V}_{X}$, so that the discrepancy
measured by 
$\norm{V_{X}\psi - \wt{V}_{X}\psi}_{2}$ is
small for \textit{any
orbital} $\psi$. The key observation of the adaptively compressed
exchange operator (ACE) is that the condition above, while desirable, is
not necessary to solve Hartree-Fock-like 
equations. In fact, it is sufficient to construct $\wt{V}_{X}$
such that $\norm{V_{X}\psi-\wt{V}_{X}\psi}_{2}$ is small when $\psi$ is
\textit{any occupied orbital}, which spans a subspace of strict rank $N_e$. 
In this sense, the ACE is designed to be adaptive to the occupied orbitals. 
When self-consistency of the occupied orbitals is reached, the physical
quantities computed in the ACE formulation is exactly the same as
that obtained with standard methods for solving Hartree-Fock-like
equations.

More specifically, in each outer iteration, for a 
given set of orbitals $\{\varphi_{i}\}_{i=1}^{N_{e}}$, we first compute
\begin{equation}
  W_{i}(\vr) = (V_{X}[\{\varphi\}]\varphi_{i})(\vr), \quad
  i=1,\ldots,N_{e}.
  \label{eqn:W}
\end{equation}
The adaptively compressed exchange operator, denoted by $\vace$, should satisfy the conditions
\begin{equation}
  (\vace \varphi_{i})(\vr) = W_{i}(\vr), \quad
   \mbox{and} \quad
  \vace(\vr,\vr') = \vace(\vr',\vr).
  \label{eqn:ACEcond}
\end{equation}
One possible choice to satisfy both conditions in
Eq.~\eqref{eqn:ACEcond} is
\begin{equation}
  \vace(\vr,\vr') = \sum_{i,j=1}^{N_{e}} W_{i}(\vr)
  B_{ij} W_{j}(\vr'),
  \label{eqn:ACEconstruct1}
\end{equation}
where $B$ is a negative semidefinite matrix to be determined, since $V_{X}$ is a negative semidefinite operator.
In order to determine
the matrix $B$, for any $k,l=1,\ldots,N_{e}$, we require
\begin{multline}
  \iint \varphi_{k}(\vr) \vace(\vr,\vr') \varphi_{l}(\vr') \ud \vr
  \ud \vr' \equiv \int \varphi_{k}(\vr) W_{l}(\vr) \ud \vr\\
  = \sum_{i,j=1}^{N_{e}} \left(\int \varphi_{k}(\vr) W_{i}(\vr) \ud
  \vr\right)
  B_{ij} \left(\int W_{j}(\vr')\varphi_{l}(\vr') \ud \vr'\right).
  \label{eqn:ACEconstruct2}
\end{multline}
Define 
$M_{kl} = \int \varphi_{k}(\vr) W_{l}(\vr) \ud \vr$,
then by Eq.~\eqref{eqn:W}, $M$ is a negative semidefinite matrix of size
$N_{e}$.
Eq.~\eqref{eqn:ACEconstruct2} can be simplified using matrix notation as
\[
M = M B M.
\]
Perform Cholesky factorization for $-M$, i.e. $M=-LL^{T}$, where $L$ is a lower
triangular matrix, then the solution to~\eqref{eqn:ACEconstruct2} is
$B=-L^{-T}L^{-1}$.
Define the projection vector in the ACE formulation as
\begin{equation}
  \xi_{k}(\vr) = \sum_{i=1}^{N_{e}} W_{i}(\vr) (L^{-T})_{ik},
  \label{eqn:Wtilde}
\end{equation}
then the adaptively compressed exchange operator is given by
\begin{equation}
  \vace(\vr,\vr') = -\sum_{k=1}^{N_{e}} \xi_{k}(\vr)
  \xi_{k}(\vr').
  \label{eqn:ACE}
\end{equation}


It should be noted that $\vace$ is an operator of strict rank
$N_{e}$. By construction $\vace$ only agrees with $V_{X}$ when
applied to $\{\varphi_{i}\}_{i=1}^{N_{e}}$. In the subspace
orthogonal to the subspace spanned by
$\{\varphi_{i}\}_{i=1}^{N_{\varphi}}$, the discrepancy between
$V_{X}$ and $\vace$ is in principle not controlled. Nonetheless, the ACE
formulation is sufficient to provide correct eigenvalues $\{\varepsilon_{i}\}$ in
Eq.~\eqref{eqn:HF} when self-consistency of the orbitals is reached. 

The main advantage of the ACE formulation is the significantly reduced
cost of applying $\vace$ to a set of orbitals than that of applying $V_X$.
Once ACE is constructed, the cost of applying $\vace$
to any orbital $\psi$
is similar to the application of a nonlocal pseudopotential,
thanks to its low rank structure.  
ACE only needs to be constructed once when $\varphi_i$'s are updated in
the outer iteration. After constructed,
the ACE can be reused for all the subsequent inner SCF iterations for the electron
density, and each iterative step for solving the linear eigenvalue
problem. Since each outer iteration could require $10\sim 100$ or more
applications of the Hamiltonian matrix $H$, the cost associated with the
solution of the Poisson problem is hence greatly reduced. The pseudocode
for iterative methods with the ACE formulation is given in
Alg.~\ref{alg:ACE}. Comparing with Alg.~\ref{alg:fock}, the ACE
formulation only requires moderate modification of the code.

We also remark that ACE can be readily used to  reduce the computational cost of the exchange energy, without the need of solving any extra Poisson equation:
\begin{equation}
E_X^{HF} = \frac{1}{2} \sum_{i=1}^{N_e} \iint
  \psi_{i}(\vr)\vace(\vr,\vr')\psi_i(\vr') \ud \vr \ud \vr' 
  =-\frac{1}{2} \sum_{i,k=1}^{N_e} \left(\int \psi_i(\vr) \xi_k(\vr) \ud \vr\right)^2.  
\label{eqn:EX_ACE}
\end{equation}

\begin{algorithm}[htbp]
\begin{center}
  \begin{minipage}{5in}
\begin{algorithmic}[1]
  \WHILE{exchange energy is not converged}
  \STATE Compute $\{W_{i}\}$ according to ~\eqref{eqn:W}.
  \STATE Compute $\{\xi_{k}\}$ according to ~\eqref{eqn:Wtilde}.
  \WHILE{electron density $\rho$ is not converged}
  \STATE Solve the linear eigenvalue problem~\eqref{eqn:HFlineig} with
  iterative schemes, with $V_{X}$ replaced by $\vace$ according to~\eqref{eqn:ACE}.
  \STATE Update $\rho^{out}(\vr)\gets \sum_{i=1}^{N_e}
  \abs{{\psi}_{i}(\vr)}^2$.
  \STATE Update $\rho$ using $\rho^{out}$ and possibly previous history of $\rho$ with charge mixing schemes.
  \ENDWHILE
  \STATE Compute the exchange energy $E_{X}$ according to~\eqref{eqn:EX_ACE}.
  \STATE Update $\{\varphi_{j}\}_{j=1}^{N_{e}}\gets
  \{\psi_{j}\}_{j=1}^{N_{e}}$.
  \ENDWHILE
\end{algorithmic}
\end{minipage}
\end{center}
\caption{Iterative methods for solving Hartree-Fock-like
equations in the ACE formulation.}
\label{alg:ACE}
\end{algorithm}

So far we assumed the number of $\{\varphi_i\}$ orbitals, denoted by
$N_{\varphi}$, is exactly equal to $N_e$. When unoccupied states are
needed, e.g. for the computation of the HOMO-LUMO gap or for excited state calculations, 
$N_{\varphi}>N_e$ should be used. We define the oversampling ratio
$r=N_{\varphi}/N_{e}$. Choosing the oversampling ratio $r>1$
can be potentially advantageous in the ACE formulation to accelerate the
convergence of the outer SCF iteration. This is because when $r>1$,  $\vace$ agrees with the true exchange operator $V_X$ when
applied to orbitals over a
larger subspace. Our numerical results, while validating this intuitive
understanding, also indicates that
the choice $r=1$ (i.e. $N_{\varphi}=N_e$) can be good enough for
practical hybrid functional calculations.

\section{Numerical results}\label{sec:numer}

In this section we demonstrate the effectiveness of the ACE formulation
for accelerating KSDFT calculations with hybrid functionals. The ACE
formulation is implemented in the
DGDFT software package~\cite{LinLuYingE2012,HuLinYang2015a}. DGDFT is a
massively parallel electronic structure software package for ground
state calculations written in C++.
It includes a relatively self-contained module called PWDFT for
performing standard planewave based electronic structure calculations.
We implement the Heyd-Scuseria-Ernzerhof
(HSE06)~\cite{HeydScuseriaErnzerhof2003,HeydScuseriaErnzerhof2006}
hybrid functional in PWDFT, using periodic boundary conditions with
$\Gamma$-point Brillouin zone sampling. The screening parameter in the HSE functional $\mu$ is set to $0.106$ au. Our implementation is comparable to that in
standard planewave based software packages such as Quantum
ESPRESSO~\cite{GiannozziBaroniBoniniEtAl2009}.  All results are performed on a single
computational core of a 3.4 GHz Intel i-7 processor with 64 GB memory.

\begin{figure}[ht]
  \begin{center}
    \subfloat[(a)]{\includegraphics[width=0.3\textwidth]{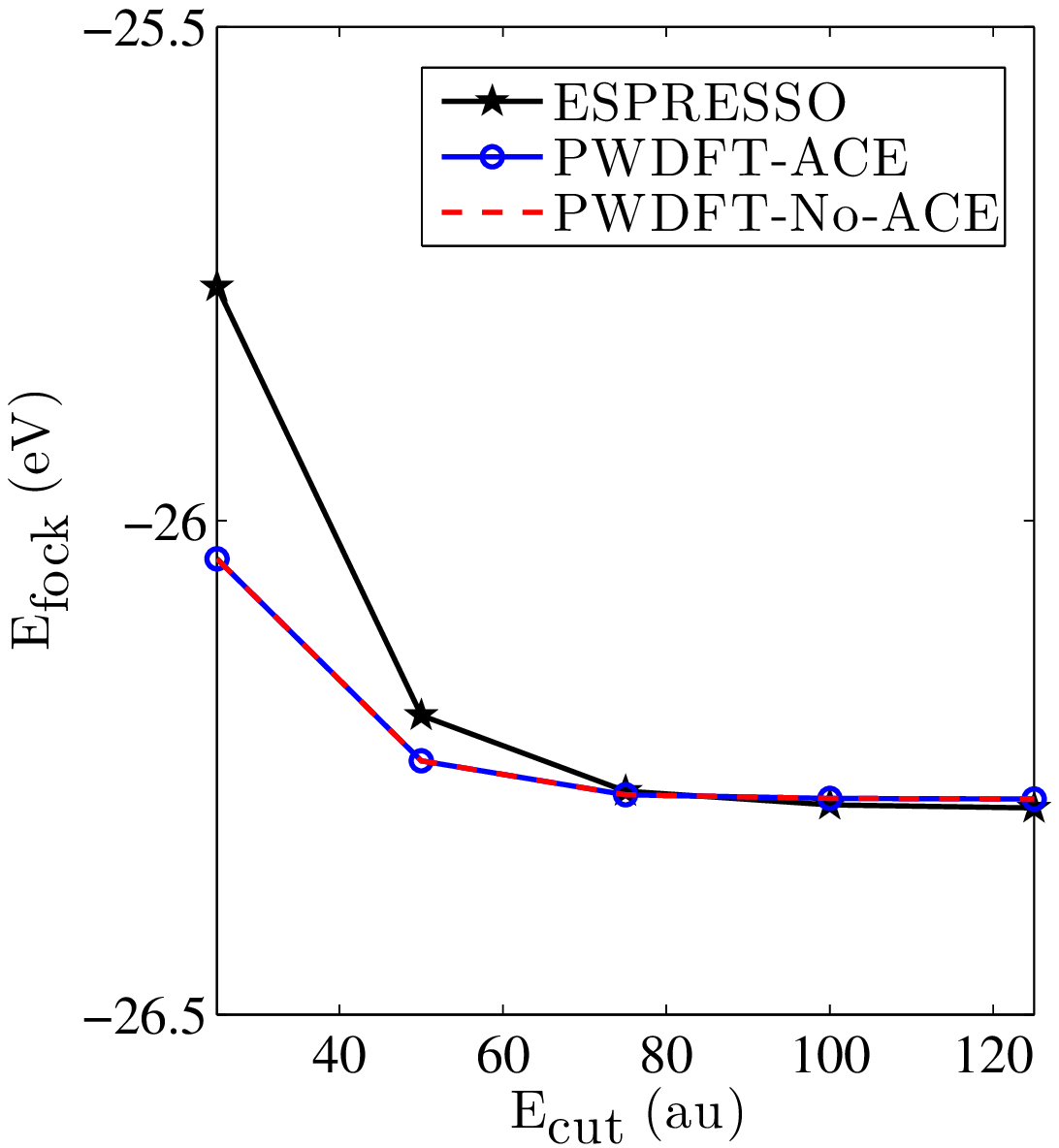}}
    \quad
    \subfloat[(b)]{\includegraphics[width=0.3\textwidth]{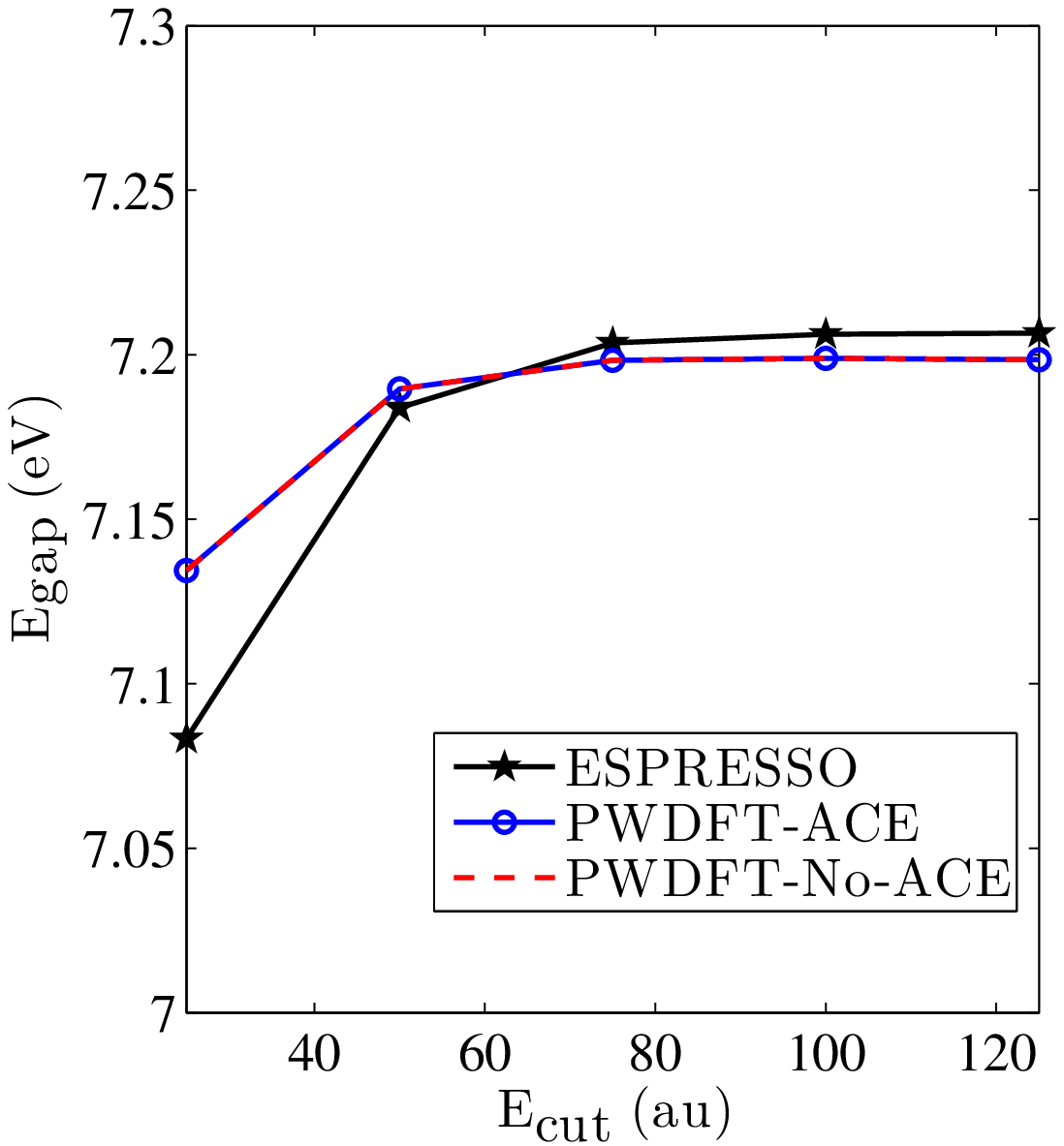}}
  \end{center}
  \caption{(color online) (a) Fock exchange energy and (b) HOMO-LUMO gap obtained from
  Quantum ESPRESSO, PWDFT with and without the ACE formulation for a
  water molecule.}
  \label{fig:water1}
\end{figure}

\begin{figure}[ht]
  \begin{center}
    \subfloat[(a)]{\includegraphics[width=0.3\textwidth]{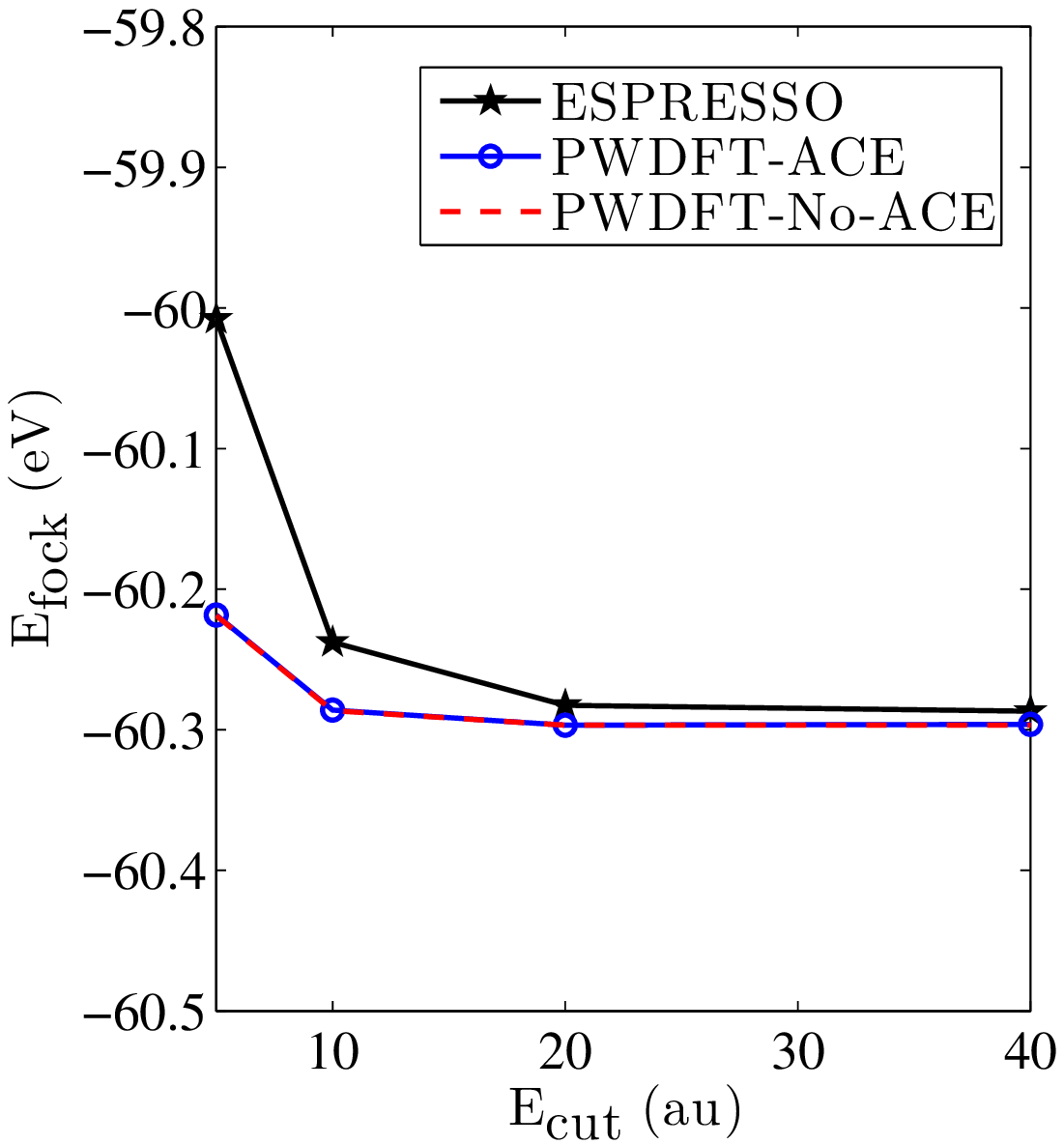}}
    \quad
    \subfloat[(b)]{\includegraphics[width=0.3\textwidth]{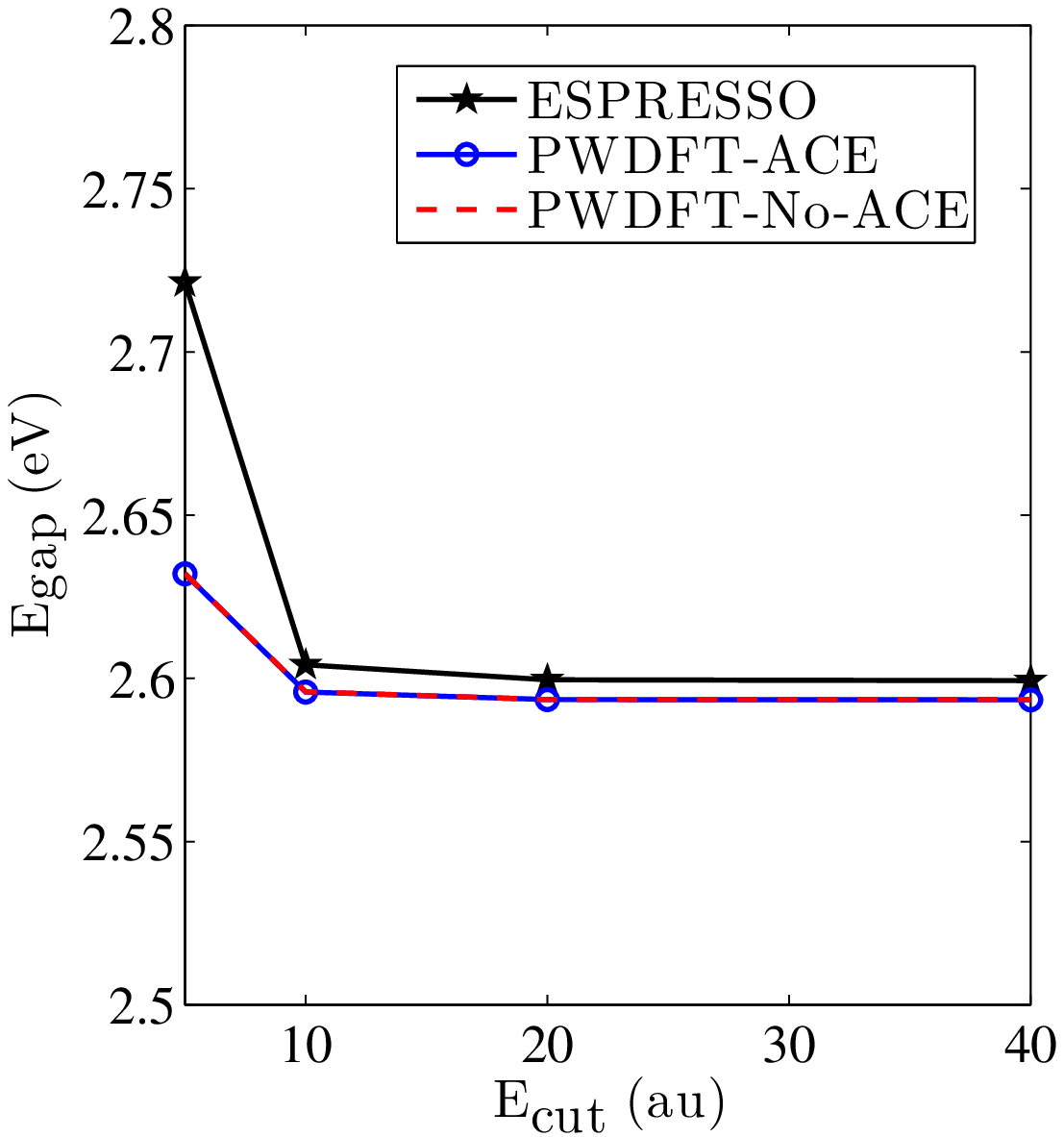}}
  \end{center}
  \caption{(color online) (a) Fock exchange energy and (b) HOMO-LUMO gap obtained from
  Quantum ESPRESSO, PWDFT with and without the ACE formulation for a
  silicon system with $8$ atoms.}
  \label{fig:si1}
\end{figure}

We first validate the accuracy of the hybrid functional implementation in
PWDFT by benchmarking with Quantum ESPRESSO, and by comparing the converged Fock exchange energy and
the HOMO-LUMO gap for a single water molecule (Fig.~\ref{fig:water1})
and an $8$-atom silicon system (Fig.~\ref{fig:si1}). The
Hartwigsen-Goedecker-Hutter (HGH)
dual-space pseudopotential~\cite{HartwigsenGoedeckerHutter1998} is used in all
calculations.  Both Quantum ESPRESSO and PWDFT control the accuracy using a
single parameter $E_{\mbox{cut}}$, the kinetic energy cutoff.
However, there is notable difference in the detailed implementation. For instance, PWDFT uses a real space implementation of the
pseudopotential with a pseudo-charge formulation~\cite{PaskSterne2005},
and implements the exchange-correlation functionals via the
LibXC~\cite{MarquesOliveiraBurnus2012} library, while Quantum ESPRESSO uses a
Fourier space implementation of the HGH pseudopotential converted from
the CPMD library~\cite{HutterCurioni2005}, and uses a self-contained
implementation of exchange-correlation functionals. Nonetheless, at
sufficiently large $E_{\mbox{cut}}$, the difference of the total Fock
exchange energy between Quantum ESPRESSO and PWDFT is only $9$ meV for the water system and 
$11$ meV for the silicon system, and the difference of the gap is $8$ meV for
the water system and $5$ meV for the silicon system, respectively. In
both systems, the difference of the results from PWDFT is negligibly
small between the standard
implementation of hybrid functional (No-ACE), and with the ACE
formulation.


\begin{figure}[ht]
  \begin{center}
    \subfloat[(a) Water molecule]{\includegraphics[width=0.3\textwidth]{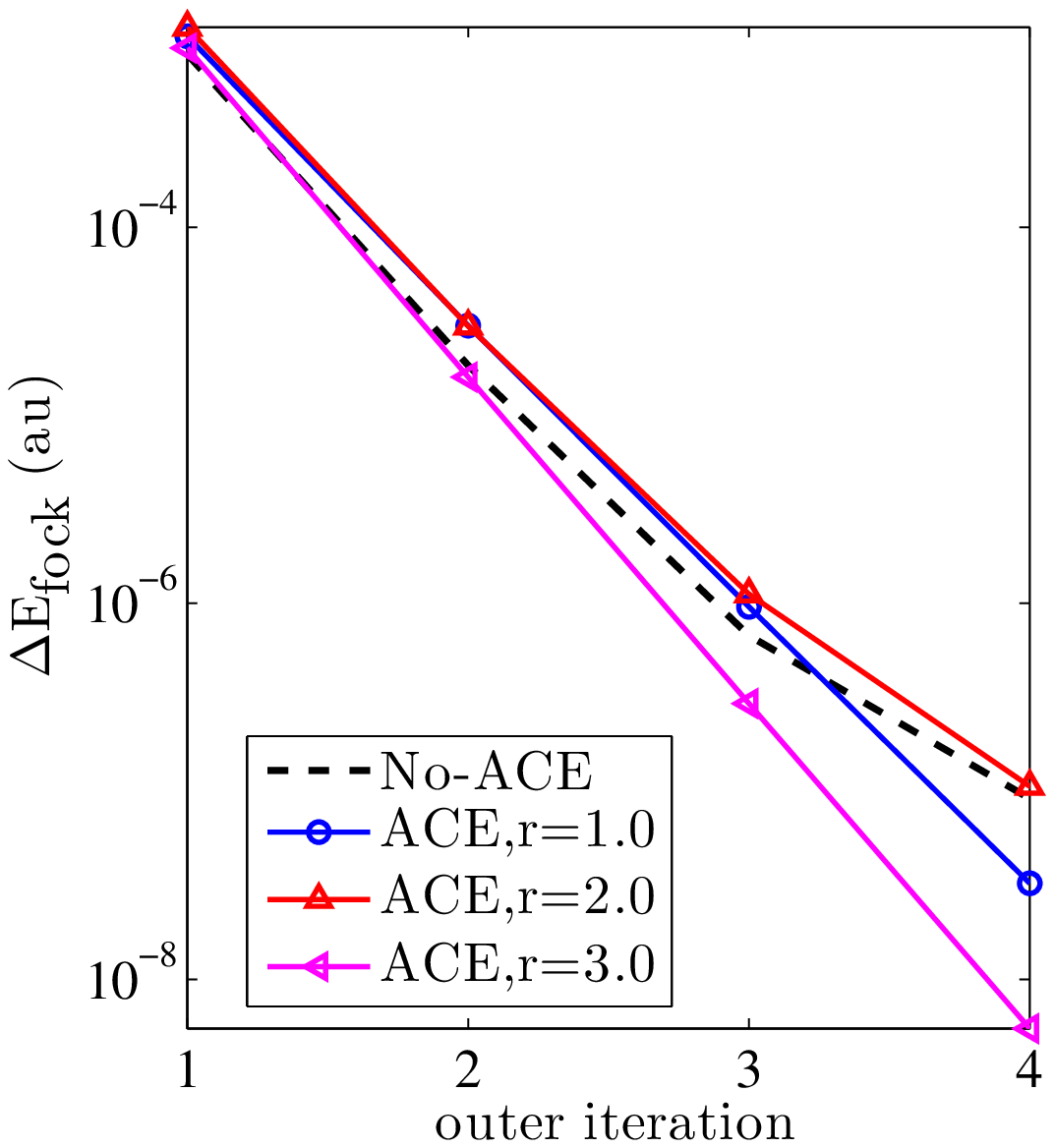}}
    \quad
    \subfloat[(b) Silicon with $8$ atoms]{\includegraphics[width=0.3\textwidth]{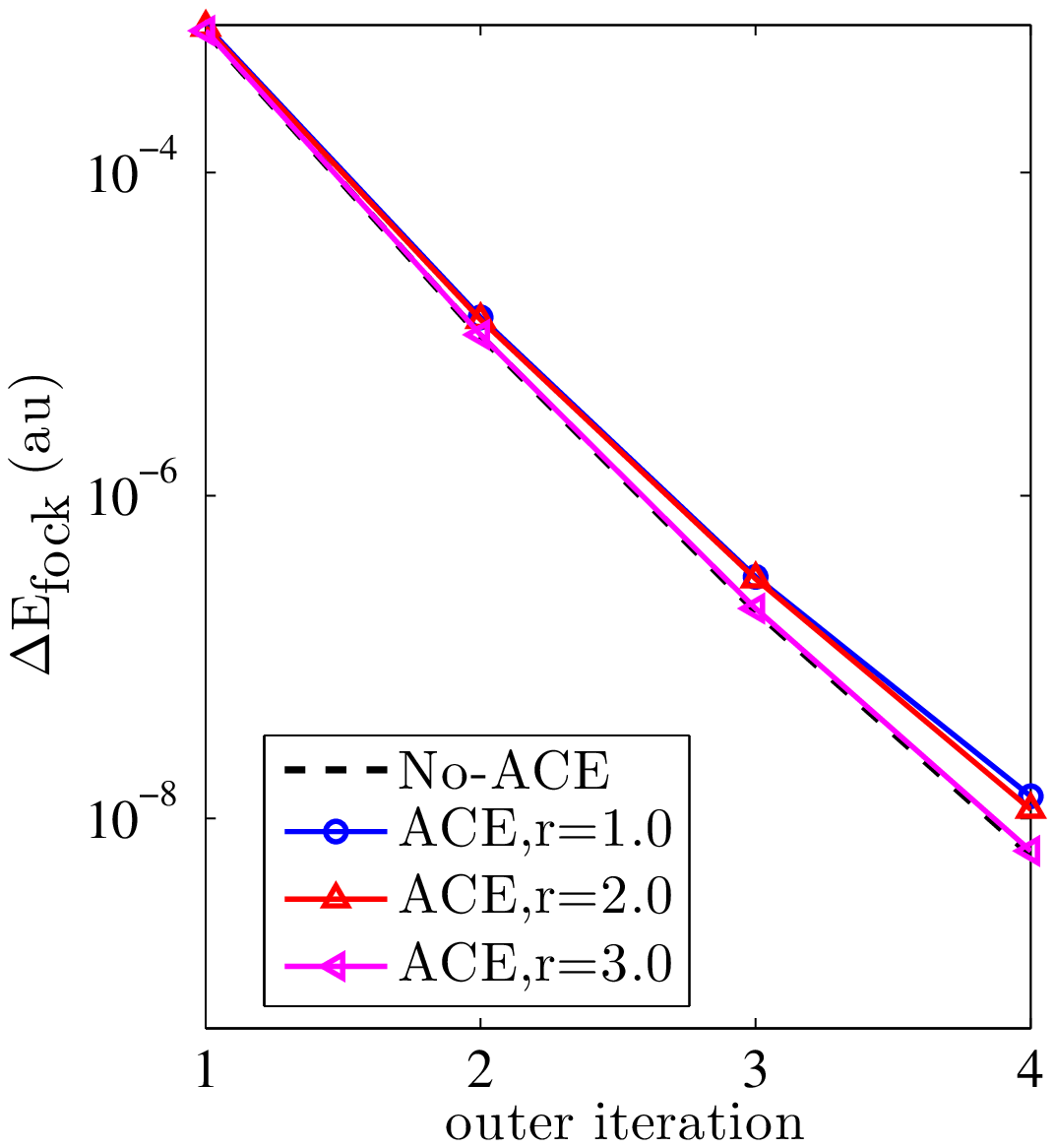}}
  \end{center}
  \caption{(color online) Convergence of the difference of the exchange energy.}
  \label{fig:oversample}
\end{figure}

In section~\ref{sec:ace} the oversampling ratio
$r=N_{\varphi}/N_{e}$ is defined. It is conceivable that as $r$
increases, the convergence of the outer iteration for the orbitals
$\{\varphi_{i}\}$ can accelerate. Fig.~\ref{fig:oversample} (a) and (b)
report the convergence of the difference of the Fock exchange energy at
each outer iteration with respect to different oversampling ratio
$r$ for the water and silicon system, respectively, as a measure of the
convergence of the outer iteration.  The kinetic energy
cutoff for the water and silicon systems is set to $100$ au and $20$ au,
respectively. The convergence without the ACE formulation is also
included fir comparison. We observe that as the oversampling
ratio increases, the convergence rate of the outer iteration becomes
marginally improved. In fact the convergence rate using the ACE
formulation with $r=1$ is very close to that without the ACE formulation
at all. This indicates that the use of the ACE formulation does not
hinder the convergence rate of the hybrid functional calculation.

\begin{figure}[ht]
  \begin{center}
    \subfloat[(a)$8$ atoms]{\includegraphics[width=0.3\textwidth]{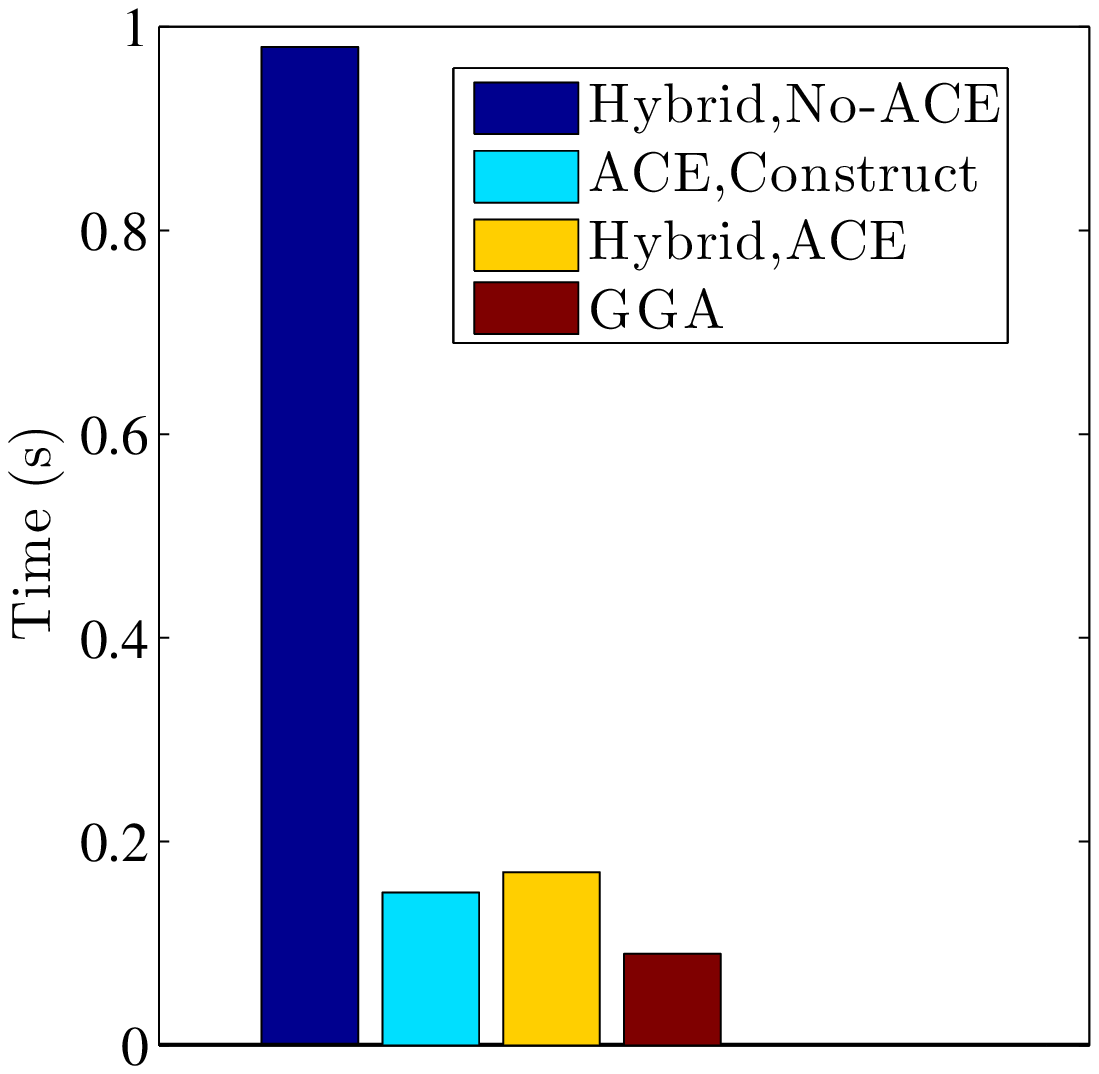}}
    \subfloat[(b)$64$ atoms]{\includegraphics[width=0.3\textwidth]{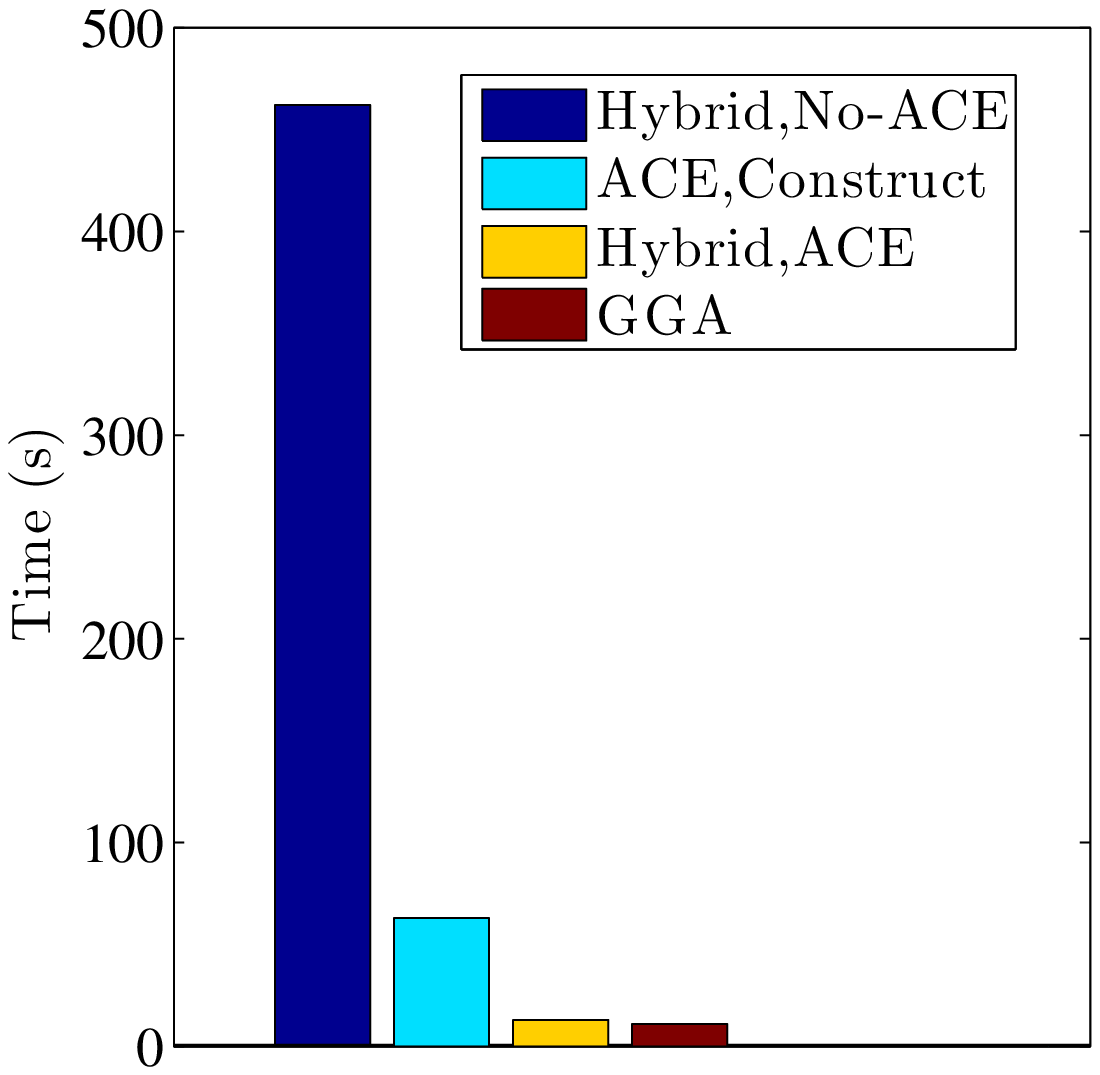}}
    \subfloat[(c)$216$ atoms]{\includegraphics[width=0.3\textwidth]{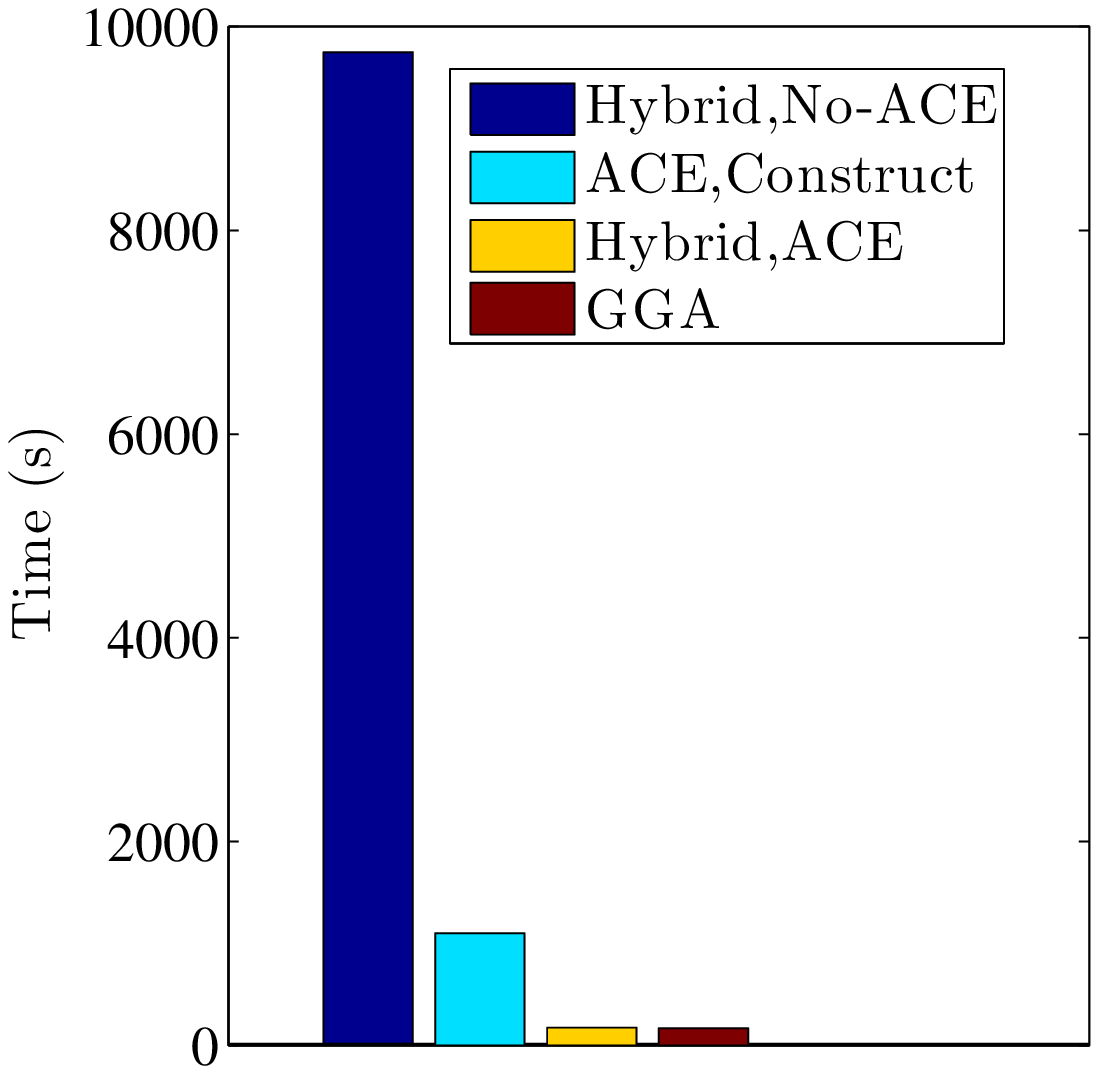}}
  \end{center}
  \caption{(color online) Computational time for silicon systems with increasing system
  sizes.}
  \label{fig:time_si}
\end{figure}

In order to demonstrate the efficiency of the ACE formulation for hybrid
functional calculations, we study three silicon systems with increasing
sizes $8$, $64$
and $216$ atoms, respectively. The latter two systems correspond to a
silicon unit cell with $8$ atoms replicated into a $2\times 2\times 2$
and a $3\times3\times 3$ supercell, respectively. Since hybrid
functional is implemented in PWDFT so far in the serial
mode, we use a
relatively small kinetic energy cutoff $E_{\mbox{cut}}=5$ au in
these calculations. Nonetheless, the kinetic energy cutoff mainly
affects the cost of the FFTs, and we expect that the ACE formulation
will become more advantageous with a higher
$E_{\mbox{cut}}$ in terms of the reduction of the absolute computational time.
Fig.~\ref{fig:time_si} shows the time cost of each SCF iteration for the
electron density, which
involves $10$ LOBPCG iterations, for the
calculation with the HSE functional with and without the ACE formulation. For comparison we also include the time cost of each SCF
iteration for the electron density in a GGA functional calculation using the
Perdew-Burke-Ernzerhof (PBE) functional~\cite{PerdewBurkeErnzerhof1996},
of which the cost is much less expensive. The cost of the construction phase of the ACE formulation is
marked separately as ``ACE,Construct'' in Fig.~\ref{fig:time_si}. 

First we confirm that the cost of each hybrid functional calculations is
much higher than that of GGA calculations. The time per SCF iteration for the
electron density of the HSE
calculation without ACE is $42$ times higher than that of the PBE calculation
for the $64$ atom system. This ratio becomes $58$ times for the $216$ atom system.
With the ACE formulation, this ratio is reduced to $1.18$ and $1.05$,
for the $64$ and $216$ atom systems, respectively, i.e. the cost of each
HSE calculation in the ACE formulation is only marginally larger than
that of the GGA calculation. Although the construction of the ACE still
requires solving a large number of Poisson equations, the overall time
is greatly reduced since the
ACE, once constructed, can be used for multiple SCFs for converging the
electron density, until the orbitals $\varphi_{i}$'s are changed in the
outer iteration. Even assuming the inner iteration only consists of
one SCF iteration, for the system with $216$ atoms, the ACE formulation
already achieves a speedup $8.8$ times compared to the standard
implementation. The ACE formulation becomes orders of magnitude more efficient when
multiple inner SCF iterations is required, which is usually the case
both in PWDFT and in other software packages such as Quantum ESPRESSO.

\section{Conclusion}\label{sec:conclusion}

We have introduced the adaptively compressed exchange operator (ACE)
formulation for compressing the Fock exchange operator. The main
advantage of the ACE formulation is that there is no loss of accuracy,
and its effectiveness does not depend on the size of the band gap. Hence
the ACE formulation can be used for insulators, semiconductors as well
as metals. We demonstrated the use of the ACE formulation in an
iterative framework for solving Hartree-Fock equations and Kohn-Sham
equations with hybrid exchange-correlation functionals.  The ACE
formulation only requires moderate modification of the code, and can
potentially be applied to all electronic structure software packages for
treating the exchange interaction.  The construction cost of the ACE
formulation is the same as applying the Fock exchange operator once to
the occupied orbitals. Once constructed, the cost of each
self-consistent field iteration for the electron density in hybrid functional calculations
becomes only marginally larger than that of  GGA calculations.
Our numerical results indicate that the computational advantage of the ACE
formulation can be clearly observed even for small systems with tens of atoms.

For insulating systems, the cost of the ACE formulation can be further
reduced when combined with linear scaling type methods. For range separated
hybrid functionals, it might even be possible to localize the projection
vectors $\xi_k$'s due to the screened Coulomb interaction in the real
space. This could further reduce the construction as well as the
application cost of the ACE, and opens the door to Hartree-Fock-like
calculations for a large range of systems beyond reach at present.

\begin{acknowledgement}
This work was partially supported by Laboratory Directed Research and
Development (LDRD) funding from Berkeley Lab, provided by the Director,
Office of Science, of the U.S. Department of Energy under Contract No.
DE-AC02-05CH11231, by the Scientific Discovery through Advanced Computing
(SciDAC) program and the Center for Applied Mathematics for Energy
Research Applications (CAMERA) funded by U.S. Department of Energy,
Office of Science, Advanced Scientific Computing Research and Basic
Energy Sciences, and by the Alfred P. Sloan fellowship.
\end{acknowledgement}

\providecommand{\latin}[1]{#1}
\providecommand*\mcitethebibliography{\thebibliography}
\csname @ifundefined\endcsname{endmcitethebibliography}
  {\let\endmcitethebibliography\endthebibliography}{}


%
%
%
%
%
%
%
%

\end{document}